\documentclass[twocolumn]{aastex62}

\shorttitle{CLOUDY view of the warm corona}
\shortauthors{Panda et al.}

\begin{document}

\title{CLOUDY view of the warm corona}

\correspondingauthor{Swayamtrupta Panda}
\email{spanda@camk.edu.pl}

\author{Swayamtrupta Panda}
\affiliation{Center for Theoretical Physics (PAN), Al. Lotnik{\'o}w 32/46, 02-668 Warsaw, Poland}
\affiliation{Nicolaus Copernicus Astronomical Center (PAN), ul. Bartycka 18, 00-716 Warsaw, Poland}

\author{Bo{\.z}ena Czerny}
\affiliation{Center for Theoretical Physics (PAN), Al. Lotnik{\'o}w 32/46, 02-668 Warsaw, Poland}

\author{Chris Done}
\affiliation{Department of Physics, University of Durham, South Road, Durham DH1 3LE, UK}

\author{Aya Kubota}
\affiliation{Department of Electronic Information Systems, Shibaura Institute of Technology, 307 Fukasaku, Minuma-ku, Saitama-shi, Saitama 337-8570, Japan}
\affiliation{Department of Physics, University of Durham, South Road, Durham DH1 3LE, UK}


\begin{abstract}
Bright active galaxies show a range of properties but many of these properties are correlated which has led to the concept of the Quasar Main Sequence. We test whether our current understanding of the quasar structure allows to reproduce the pattern observed in the optical plane formed by the kinematic line width of H$\beta$ and the relative importance of the Fe II  optical emission. We performed simulations of the H$\beta$ and Fe II production using the code CLOUDY and well justified assumptions about the broad band spectra, distance of the emission line region, and the cloud properties. We show that the presence of the warm corona is an important element of the broad band spectrum which decreases the dependence of the relative Fe II emissivity on the Eddington ratio, and allows to reproduce the rare cases of the particularly strong Fe II emitters. Results are sensitive to the adopted cloud distance, and strong Fe II emission can be obtain either by adopting strongly super-solar metallicity, or much shorter distance than traditionally obtained from reverberation mapping.  We modeled in a similar way the UV plane defined by the Mg II line and Fe II UV pseudo-continuum, but here our approach is less successful, in general overproducing the Fe II strength. We found that the Fe II optical and UV emissivity depend in a different way on the turbulent velocity and metallicity, and the best extension of the model in order to cover both planes is to allow very large turbulent velocities in the Broad Line Region clouds.

\end{abstract}



\section{Introduction} \label{sec:intro}

Active Galactic Nuclei (AGN) are complex systems with properties dependent on the central black hole as well as on the surrounding medium. Unification picture leads to division of sources in Type 1 and Type 2, depending on the orientation of the observer with respect to the symmetry axis (for a review , see \citealt{netzer2015}) and their abilities to produce a strong jet (for a review , see \citealt{padovani2017}). However, even if we concentrate on Type 1 AGN without strong jets, where the central parts are not shielded from our view and the Doppler-boosted jet does not contribute to the broad band spectrum, we observe a broad range of the nucleus properties. They show as a dispersion in the  measured emission lines intensities and kinematic width, absolute luminosities, and the broad band indices. 
\par
Measurements of numerous properties in each quasar called for a search of some pattern in these properties. The essential step was made by \cite{bg92} with the use of the Principal Component Analysis (PCA). This line of research was pursued by many authors \citep{dul96,sul00,bor02,kur09,mar14} and has lead to the concept of the Quasar Main Sequence.
In the simplest version, it can be reduced to the optical plane, when only two quantities are considered: kinematic width of the H$\beta$ line and the ratio $\mathrm{R_{Fe II}}$ of the equivalent width (EW) of the Fe II emission in the  4434-4684 \AA~ range to EW of H$\beta$ line (see e.g. \citealt{bg92,sul00,sh14,mar18}). Quasar Main Sequence forms a characteristic pattern in this optical plane. 

In our recent paper (\citealt{panda18b}), we sought to model this Quasar Main Sequence from a theoretical viewpoint. We wanted to see what are the key drivers behind this pattern. It has long been considered that the Eddington ratio (\citealt{sh14}) plays an important role, and the viewing angle was suggested to be a second key parameter, although a trend with black hole mass was also noticed \citep{sh14}. We modeled the AGN sample assuming a range of black hole masses and Eddington ratios, neglecting the issue of the viewing angle and spin, and we calculated the line widths and strengths for each source under some assumptions about the Spectral Energy Distribution (SED) shape, BLR distance, density, metallicity and turbulence in the BLR. Under these assumptions, we were able to locate our modeled quasars in the optical plane.  

We found that although Eddington ratio indeed plays a role in modeling the sequence, yet it definitely needs to be coupled with few other parameters. The cloud's density is important as well, and so is the effect of turbulence within the cloud. Also, the effect of metallicity has an important role here, especially to model these strong Fe II emitters. Solar abundances can indeed explain the low Fe II content part of the diagram, but one needs to consider super-Solar chemical composition if we aim to explain the far-right end of the main sequence diagram. We were able to covered the optical plane quite well although the viewing angle was not included. Nevertheless one problem remained: we found that a simple increase in the metallicity factor allows us to cover up to 98\% of the observed sample but is not enough to explain the most extreme Fe II emitters. So the problem is not fully solved. 
\par 
In the previous work we modeled the SED assuming a contribution from the cold Keplerian disk and a contribution from a hot corona responsible for hard X-ray emission. However, observed AGN spectra usually contain another spectral component, observationally described as ``soft X-ray excess'' \citep{arnaud1985}. This component helps to bridge the absorption gap between the UV downturn and the soft X-ray upturn (\citealt{elvis94}; \citealt{laor97}; \citealt{rich06}). This component is particularly strong in Narrow Line Seyfert 1 galaxies, but may also carry a dominant fraction of the luminosity in the SED of AGN at lower Eddington ratios (\citealt{jin12a,jin12b}). Theoretically, this component is modeled as a warm corona with the temperature of the order of 1 keV \citep{mag1998,czerny2003,gier2004,porquet2004,petrucci2013,middei2018,porquet2018,petrucci2018,middei2019} which covers part of the disk and comptonizes disk photons. This component changes the far-UV and soft X-ray part of the spectrum and thus can affect the Fe II line production. 

In this paper we use a new complete model of the AGN SED  (\citealt{kub18}), which accounts for an outer standard disk, a hot corona and an inner warm Comptonizing region to produce the soft X-ray excess. We test the role of the warm corona in shaping Quasar Main Sequence and we aim at explaining the presence of the strongest Fe II emitters.

\begin{figure}
    \centering
    \includegraphics[scale=0.5]{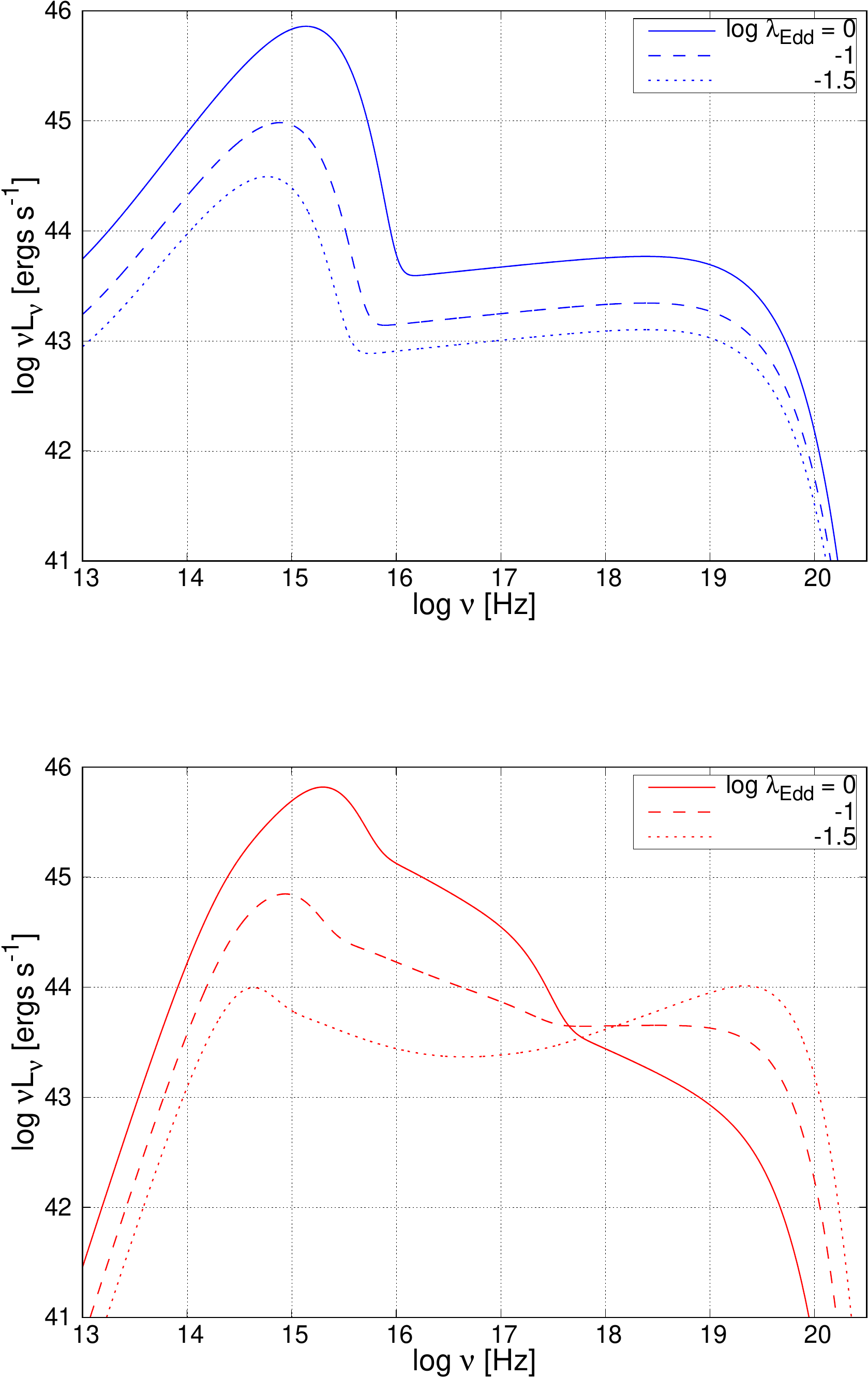}
    \caption{Comparison of the SED models from the \citealt{panda18b} (P18; upper panel) and \citealt{kub18} (WC; lower panel) for an exemplary case, M$_{\mathrm{BH}}$=6.31$\times 10^9\; \mathrm{M_{\odot}}$. The SEDs are shown for three cases of $\lambda_{\mathrm{Edd}}$.} 
    \label{fig:SED}
\end{figure}

\section{Model}
\label{sect:model}
The SED model of \cite{kub18} assumes that the accretion flow is completely radially stratified and emits as a standard disk blackbody from R$_{\mathrm{out}}$ to R$_{\mathrm{warm}}$, as warm Comptonization from R$_{\mathrm{warm}}$ to R$_{\mathrm{hot}}$, and eventually makes a transition to the hot corona part from R$_{\mathrm{hot}}$ to R$_{\mathrm{ISCO}}$. We use the subset of the SED models which are a function of black hole mass, $M$, and the Eddington ratio only. 

The simplifications of the original model are the following: The black hole spin here is fixed at \textit{a} = 0 (non-rotating black hole). The fraction of the energy dissipated in the hot corona, $L_{diss,hot}$, is fixed at 0.02$\rm{L_{Edd}}$. This $L_{diss,hot}$ and $kT_{e,hot}$ = 100 keV defines the radius of the hot corona, r$_{hot}$ = 23 for $\dot{m}$ = 0.05.  The radius of the warm corona is at twice the radius of the hot corona as per their simplified QSOSED version. The motivation for such an assumption is discussed in detail \cite{kub18} and shown in Figure 3 in their paper. We tested the effect of change in $L_{diss,hot}$ on the values of the $\mathrm{R_{FeII}}$ obtained. $\mathrm{R_{FeII}}$ increased for all the Eddington ratios -- from a meagre 1\% at $\lambda_{\mathrm{Edd}}$ = 1, to almost 40\% for the lowest values of Eddington ratio considered ($\lambda_{\mathrm{Edd}}$ = 0.03). Although not much is changed in the SED in the optical-UV part, but a change in the $L_{diss,hot}$ affects significantly the soft and hard X-ray component, enhancing the overall energy being dissipated from the coronal part affecting the higher excitation levels of FeII, which increases the total FeII strength.

These assumptions allow to get the complete SED. The disk component is modeled as  Novikov-Thorne \citep{nt73} blackbody spectrum that is modified accounting for electron scattering. This has been approximated using a colour-temperature corrected blackbody spectrum. The color-temperature correction is important especially when close to the hydrogen ionization at $\sim$ $10^4$ K. This eventually shifts the peak of the resulting blackbody spectrum rightwards by a factor $f_{col}$ \citep{kub18} which can be seen in Figure \ref{fig:SED}. This Big Blue Bump shift towards higher temperatures decreases as the Eddington ratios goes down. Since the net flux remains the same, the disk component in optical-UV has now a relatively lower normalization. This can be seen in Figure \ref{fig:SED}. The optical depth for the warm corona component in their model is defined by the spectral index of the comptonisation ($\Gamma_{e, warm}$=2.5) for an electron temperature $T_e$ = 0.2 keV (see Table 3 in \citealt{kub18}). Finally, The inclination angle used in our models is fixed at $45^o$ (for Type 1 AGNs, $i\;\epsilon\;[0^o,60^o]$; see \cite{marin14} and references therein). 

With this approach, we construct a grid of models in mass ($M_{BH}\;=\;10^6 - 10^{10}\;M_{\odot}$), and in Eddington ratio ($\lambda_{Edd}\;=\;0.03 - 1 $).  This range is consistent with the observed range for 545 SDSS quasars from \cite{lusso17}.

The remaining part of the modelling is done basically in the same way as in \citet{panda18b} although in this approach we do not need to use UV/X-ray scaling law of \cite{lusso17}. We assume that the BLR radius is given by the \citet{bentz13} law (but we also discuss the possible consequences of the departure from this law in Sec. \ref{subsect:rblr}), we consider the same density for all clouds, we include turbulence and a range of metallicities, and the computations are done using the CLOUDY code, version 17.01 (\citealt{f17}). 

\section{Results}
We use the model of the emission line production to test whether our current understanding of the BLR clouds allows us to reproduce the Quasar Main Sequence pattern in the optical plane. In particular, we study the effect of the warm corona on the BLR using the SEDs from \citet{kub18}. We compare the results to those obtained by \citet{panda18b} where such component was not included.

\subsection{Comparison of the two SED models}
In Figure \ref{fig:SED}, we show the distinctive change in the SED shape between the two models - the warm corona (hereafter WC) and the two-component standard model (\citealt{panda18b}; hereafter P18). Here we consider the spectra (for an exemplary black hole mass M$_{\mathrm{BH}}$=6.31$\times 10^9\; \mathrm{M_{\odot}}$) as a function of Eddington ratio (at $\lambda_{\mathrm{Edd}}$ = 0.03, 0.1 \& 1).  We can clearly see the warm corona component standing out in the case of Eddington limit, and its effect lessens with the drop in the Eddington ratio. The effect of the color-temperature correction in the WC models can also be seen clearly (the disk component is shifted toward higher frequencies) as opposed to the P18 models where this effect wasn't accounted for. 

\subsection{The effect of the warm corona on Fe II production}
In P18, we found that the Fe II production was increasing with the Eddington ratio. We thus first check whether this trend is preserved when the presence of the warm corona is taken into account - decreasing the Eddington ratio yielded in an increase in the net optical Fe II strength. 
\par 
 
\begin{figure*}
    \centering
    \includegraphics[scale=0.475]{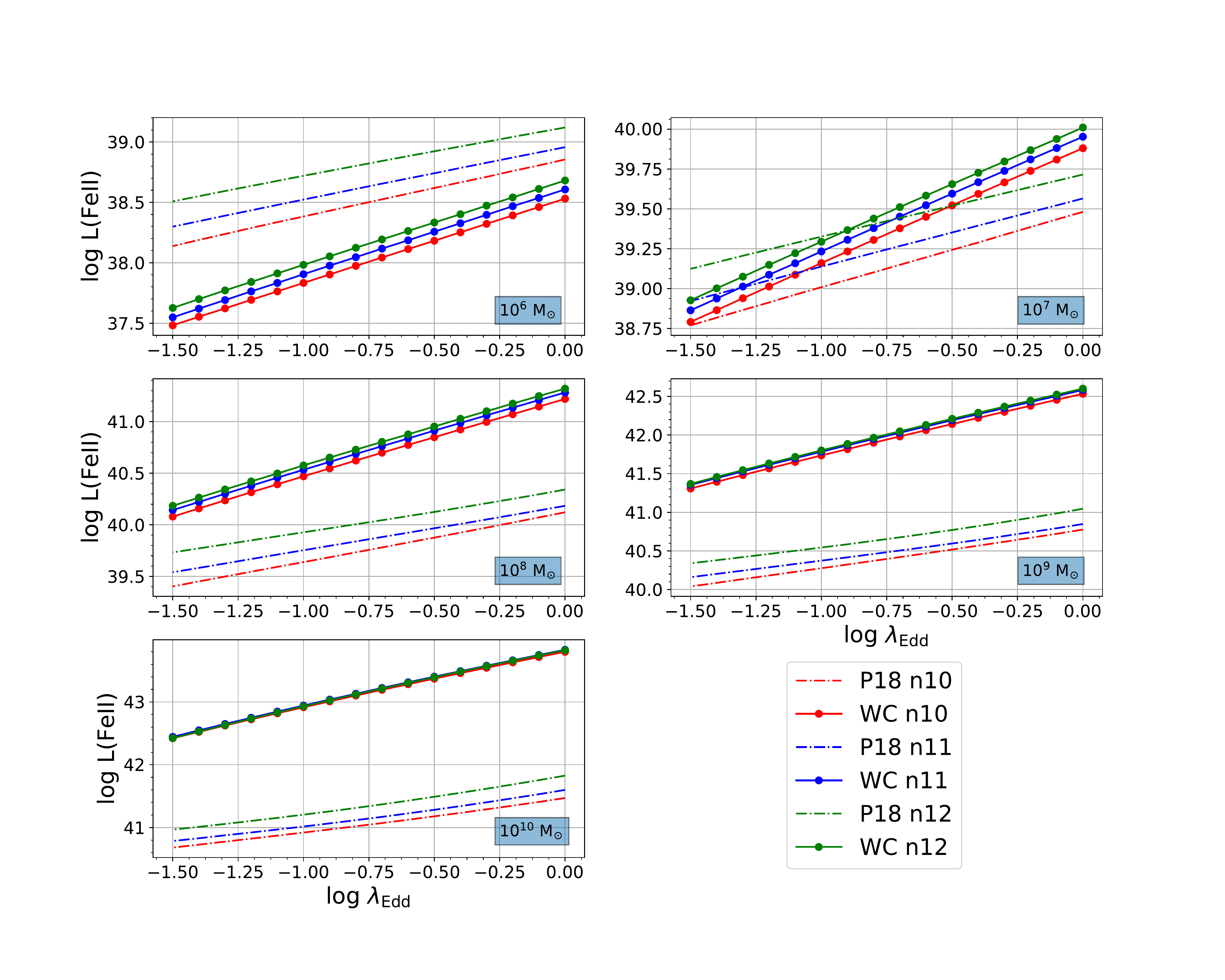}
    \caption{Optical Fe II (integrated) line luminosities vs $\lambda_{\mathrm{Edd}}$ for $\mathrm{M_{BH}} = 10^{6} - 10^{10}\; \mathrm{M_{\odot}}$ in two models: P18 and WC, assuming no turbulence and with solar abundances.}
    \label{fig:Fe II_luminosity}
\end{figure*}

In Figure \ref{fig:Fe II_luminosity} we plot the integrated Fe II (from 4434-4684 \AA, according to the \citealt{bg92} prescription which refers to the blue part of the Fe II contamination lying to the left of the H$\beta$ emission line). Models are computed for three values of cloud densities i.e., log $\mathrm{n_H}$ (in $\mathrm{cm^{-3}}$) = 10, 11, and 12, and for the specific value of hydrogen column density ($\mathrm{N_H}$ = 10$^{24}\;\mathrm{cm^{-2}}$).  Here, we show the trends for five cases of black hole masses ($\mathrm{M_{BH}} = 10^6 - 10^{10}\;\mathrm{M_{\odot}}$) that cover the full range of the models. We see that the Fe II line luminosity rises with the Eddington ratio as before, for all values of the black hole mass. However, this rise is now generally steeper that in the previous P18 models. In these computations the turbulence was not included in the models, and solar abundance was assumed.

\begin{figure*}
    \centering
    \includegraphics[scale=0.475]{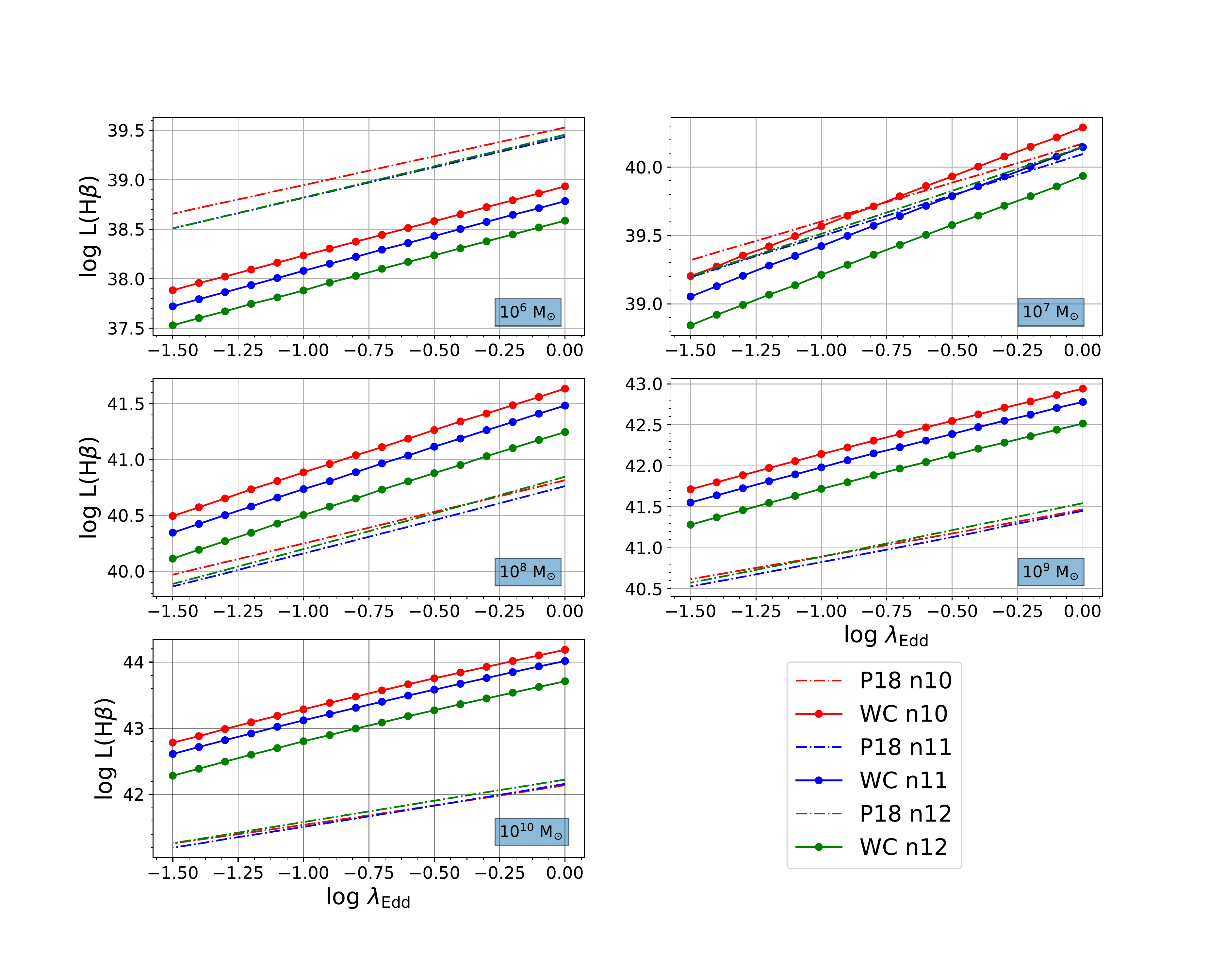}
    \caption{H$\beta$ line luminosities vs $\lambda_{\mathrm{Edd}}$ for $\mathrm{M_{BH}} = 10^{6} - 10^{10}\; \mathrm{M_{\odot}}$ in two models: P18 and WC, assuming no turbulence and with solar abundances.}
    \label{fig:Hbeta_emissivity}
\end{figure*}
\begin{figure*}
    \centering
    \includegraphics[scale=0.475]{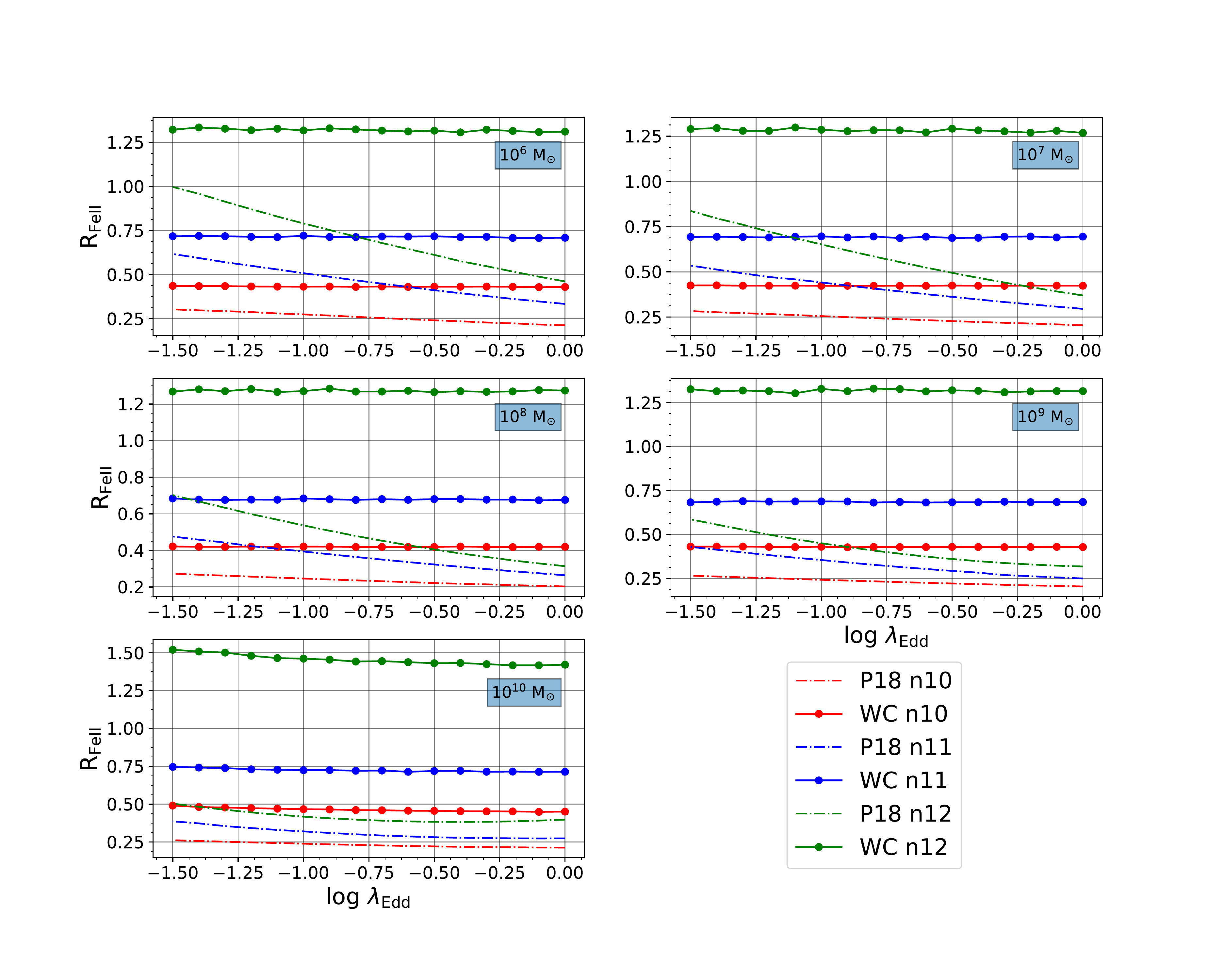}
    \caption{$\mathrm{R_{Fe II}}$ vs $\lambda_{\mathrm{Edd}}$ for $\mathrm{M_{BH}} = 10^{6} - 10^{10}\; \mathrm{M_{\odot}}$ in P18 and WC models. Almost no dependence on Eddington ratio is seen in the model with warm corona.}
    \label{fig:R_Fe}
\end{figure*}

In Figure \ref{fig:Hbeta_emissivity} we show similar plots for the  H$\beta$ line luminosities.  Again, the line luminosity rises with the Eddington ratio both in P18 model and in the present WC model. The rise again is steeper if the contribution from the warm corona is included. 

The results become more interesting when we finally plot the ratio of the Fe II and H$\beta$ (see Figure \ref{fig:R_Fe}). Without the warm corona, this value (i.e. the parameter $\mathrm{R_{Fe II}}$) showed a declining trend with increasing Eddington ratio. But with the new results for the warm corona model, the previous trend between $\mathrm{R_{Fe II}}$ and the Eddington ratio disappears. The ratio $\mathrm{R_{Fe II}}$ slightly depends on the black hole mass but it is $\sim$constant for a given mass, and for all values of the local cloud density. The values themselves are almost universal, with only a slight trend with the cloud density: higher density leads to slightly more efficient Fe II production in comparison to H$\beta$.

This will have important consequences on the trends observed in the optical plane. It was frequently argued that high values of $\mathrm{R_{Fe II}}$ correspond to high values of the Eddington ratio \citep[e.g.][]{mar18}. On the other hand, sources identified by Super-Eddington Accreting Massive Black Holes (SEAMBH) project as high Eddington ratio sources \citep{dupu2014,dupu2016,dupu2018} show quite a broad range of values of $\mathrm{R_{Fe II}}$, from 0.5 to 2 (see Fig.~3 in \citealt{czerny_NLS1_2018}). From the observational point of view, the issue is quite open.

The value of $\mathrm{R_{Fe II}}$ are now somewhat higher than obtained in P18, particularly in the case of high Eddington ratio which is promising if we aim to cover well the whole optical plane occupied by the observational points. However, here the discussion did not include option of higher metallicity and the effect of the turbulence. This we address in the next section.

\subsection{The effect of the warm corona on the quasar optical plane}

\begin{figure}
    \centering
    \includegraphics[width=8.5cm, height=3.5cm]{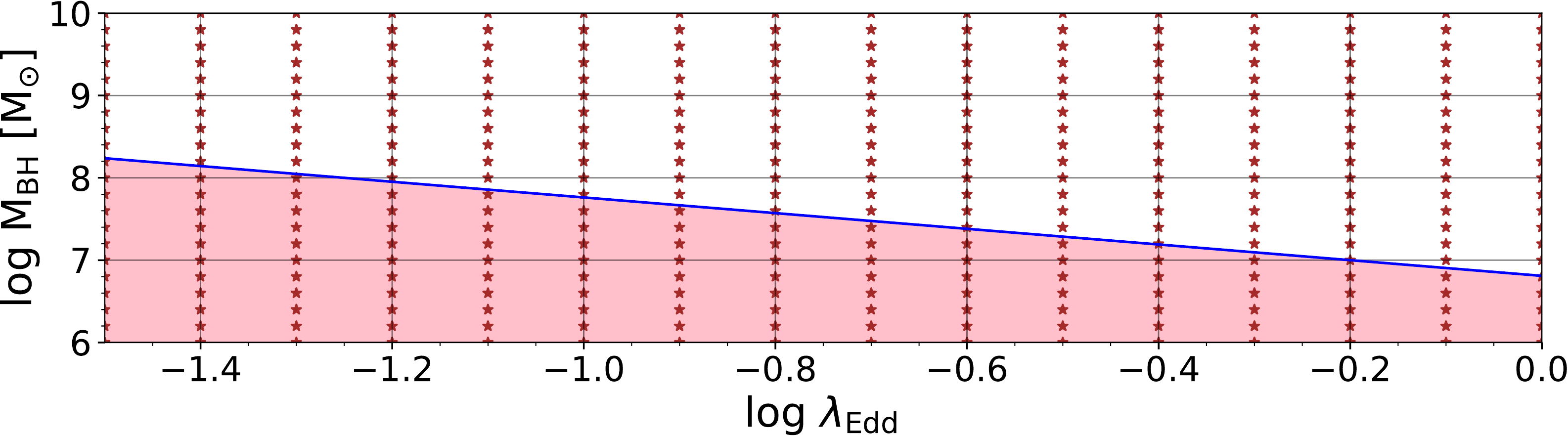}
    \caption{The parameter space for the construction of the optical plane: The pink shaded region represents the unobserved region in the \citet{shen11} quasar sample. The stars represent the grid points used in the computations.}
    \label{fig:param_space}
\end{figure}

Finally, we aim to reproduce statistically the coverage of the optical plane with our modeled sources. We model the whole range of masses and accretion rates, as described in Sect.~\ref{sect:model}. However, the observational construction of the optical plane is biased by the choice of only high Eddington ratio sources in the case of small black hole mass due to the flux limits in the quasar sample. We showed in P18 that the quasars in \citet{shen11} catalog populating the $\log \mathrm{M_{BH}}$ - $\log \lambda_{\mathrm{Edd}}$ plane are limited by the relation log($\lambda_{\mathrm{Edd}}$) = -1.05 log($\mathrm{M_{BH}}$) + 7.15. We show this in Figure~\ref{fig:param_space}, and in further study we take into consideration mostly models which populate the white part of the diagram.

\begin{figure*}
    \centering
    \includegraphics[scale=0.5]{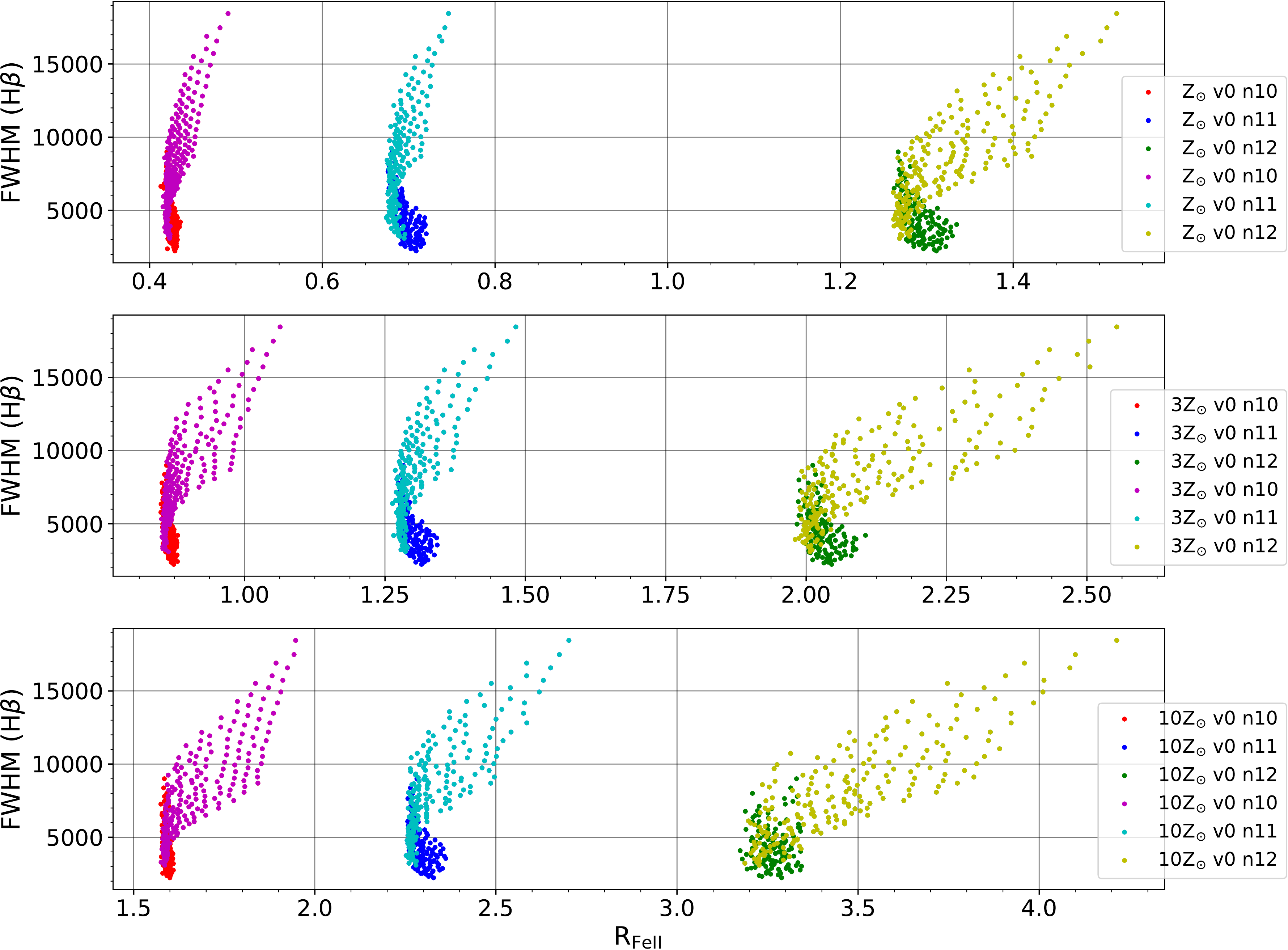}
    \caption{The coverage of the optical plane for a range of abundances (1, 3 and 10 times solar), changing cloud densities (log $\mathrm{n_H}$  (in $\mathrm{cm^{-3}}$) = 10, 11, and 12) at constant turbulent velocity (0 km/s). Two cases are shown for each value of the parameter: lighter color (magenta, light blue and light green) represent the limited parameter range shown in Fig.~\ref{fig:param_space}, and the remaining objects in  the whole mass-Eddington ratio range are shown in darker color (red, blue and green).}
    \label{fig:opt_plane_density}
\end{figure*}

\begin{figure*}
    \centering
    \includegraphics[scale=0.475]{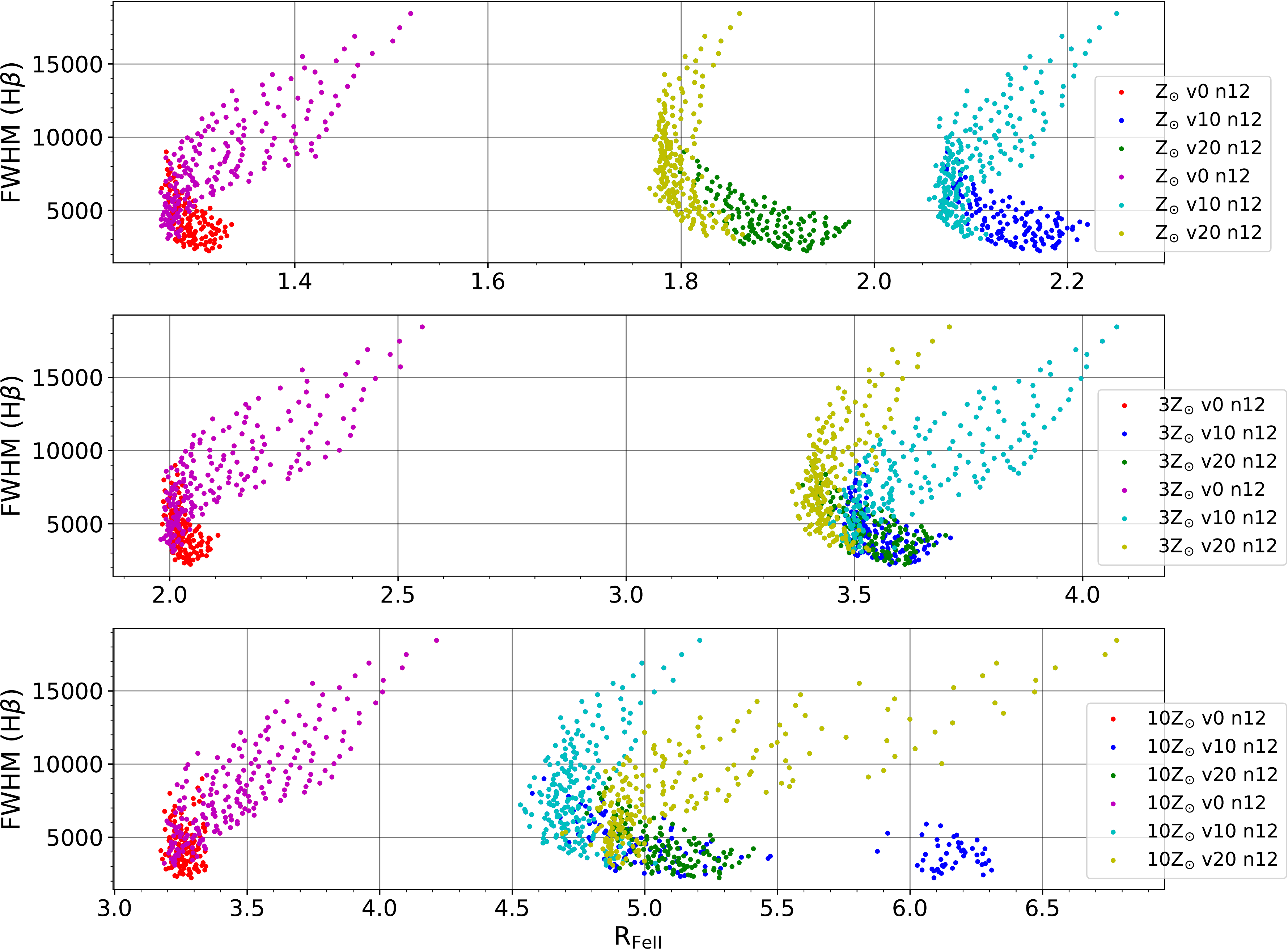}
    \caption{The coverage of the optical plane for a range of abundances (1, 3 and 10 times solar), changing turbulent velocities (0, 10 and 20 km/s) at constant cloud density ($\mathrm{n_H = 10^{12}\;cm^{-3}}$). Two cases are color-coded as in Fig.~\ref{fig:opt_plane_density}.}
    \label{fig:opt_plane_turbulence}
\end{figure*}

Figures \ref{fig:opt_plane_density} and \ref{fig:opt_plane_turbulence} show the full scale of the modeled sequence with the warm corona model. In Figure \ref{fig:opt_plane_density} the $\mathrm{v_{FWHM}}-\mathrm{R_{Fe II}}$ diagram is shown for varying cloud densities (log $\mathrm{n_H}$  (in $\mathrm{cm^{-3}}$) = 10, 11, and 12), without the effect of turbulence. The three panels show the effect of changing abundances ($\mathrm{Z=Z_{\odot},\;3Z_{\odot}\;\&\;10Z_{\odot}}$). The plots show the full sequence ($\mathrm{M_{BH}}= 10^6 - 10^{10}\;\mathrm{M_{\odot}};$ $\lambda_{\mathrm{Edd}}$ from $\sim$3\% up to Eddington limit) and the sequence constrained by the detection limit given in Figure \ref{fig:param_space}. In our analysis we derive the FWHM value from the BLR radius and black hole mass assuming the value of the virial factor 1  (see Eq.~1 in P18) for simplicity while \citet{sh14} and \citet{mejia18} show strong coupling between the virial factor and the line width.

An increase of the cloud density clearly leads to an increase in the Fe II strengths (overall rise by a factor of $\sim 1.65$ going from $\mathrm{n_H\;(in\;cm^{-3})} = 10^{10}$ to $10^{11}$; and by a factor of $\sim 3.1$ going from $\mathrm{n_H\;(in\;cm^{-3})} = 10^{10}$ to $10^{12}$). However, the predicted increase of the Fe II strength is correlated with the H$\beta$ line widths: relatively stronger Fe II emission is expected for broader line galaxies. This is rather unexpected as there aren't many Seyfert 1 galaxies detected with such high Fe II strengths. This may point out towards the dependence between the cloud density and the kinematic line width which can be achieved through an intrinsic dependence between cloud density and the Eddington ratio. Arguments for such coupling can be found for example in \citet{adhikari16}, where high densities were required to form Lorentzian line profiles without a clear gap between the BLR and Narrow Line Region (NLR), a characteristic for NLS1.

As we increase the overall cloud abundances (from solar to 10 times solar) we do obtain Fe II strengths that are comparable to those obtained for ``strong'' NLS1s sources (occupying the rightward tail region in the \cite{sh14} Quasar Main Sequence diagram). As it can be seen, in the limited sequence there is a significant number of the expected NLS1s that are excluded. There is an overall rise in $\mathrm{R_{Fe II}}$ from $\sim$ 0.42 ($\mathrm{n_H} = 10^{10}\;\mathrm{cm^{-3}};\;\mathrm{Z = Z_{\odot}}$) to $\sim$ 4.5 ($\mathrm{n_H} = 10^{12}\;\mathrm{cm^{-3}};\; \mathrm{Z = 10Z_{\odot}}$). These values for the narrower broad H$\beta$ cases range from $\leq 0.44$ ($\mathrm{n_H} = 10^{10}\;\mathrm{cm^{-3}};\;\mathrm{Z = Z_{\odot}}$) to $\leq 3.45$ ($\mathrm{n_H} = 10^{12}\;\mathrm{cm^{-3}}$; $\mathrm{Z = 10Z_{\odot}}$). 

Another interesting result is that none of the modeled values for the FWHM of the H$\beta$ line drops below the limit of 2000 km/s, so formally our current model does not cover the NLS1. Our black hole mass range includes small values, as we start from $10^6\;\mathrm{M_{\odot}}$. In the \citet{sh14} diagram there are sources with FWHM below 1500 km s$^{-1}$, although not too many. The absence of the narrow line objects in our model is likely related to two effects -- assuming the value of the virial factor 1  (see Eq.~1 in P18) for simplicity while \citet{sh14} and \citet{mejia18} show strong coupling between the virial factor and the line width. Thus virial factor for narrower lines should be considerably higher, up to a factor few, and in this way we would obtain the FMHM from the model smaller by the same factor. In the present study we decided not to use more complex prescription for the line width vs. black hole mass relation in order not to complicate the picture too much. The second plausible reason is that we do not include in the present study the problem of the viewing angle range, and the viewing angle can have some effect on the measured line width \citep{sh14,sun18}.    
\par 
In Figure \ref{fig:opt_plane_turbulence}, a similar $\mathrm{v_{FWHM}}-\mathrm{R_{Fe II}}$ diagram is shown but keeping the cloud density constant at high value ($\mathrm{n_H} = 10^{12}$ cm$^{-3}$) and changing the turbulence within the cloud ($\mathrm{v_{turb}} = 0,\;10\;\&\;20\;\mathrm{km/s}$). Similar to the Figure \ref{fig:opt_plane_density}, three cases of abundances are considered. The panels clearly show the effect of coupling between the high density within the clouds and a non-negligible microturbulence. We observe an increase in the overall Fe II strengths (starting from $\sim1.3$ for $\mathrm{v_{turb}} = 0$ km/s, $\mathrm{Z = Z_{\odot}}$; to almost a factor of 4.5 increase for the case with $\mathrm{v_{turb}} = 20$ km/s, $\mathrm{Z = 10Z_{\odot}}$). However, unlike in the Figure \ref{fig:opt_plane_density}, the almost monotonic behaviour in the $\mathrm{v_{FWHM}}-\mathrm{R_{Fe II}}$ plane is gone. Increasing the turbulence to a finite value (10-20 km/s), even for the limited sequence (see Fig.~\ref{fig:opt_plane_turbulence}), shows a marked increase in the number of low-FWHM sources, although large number of low-FWHM sources get excluded again with this constraint. As in our findings in P18, the case with 10 km/s usually gives the highest Fe II strength (compared to the other two cases). Further increase of the microturbulence leads in general to a decline in the strengths. But, increasing the abundances, makes the strengths obtained comparable in the $\mathrm{v_{turb}}$=10 km/s and 20 km/s cases - for the $\mathrm{Z = 10Z_{\odot}}$ case, the higher turbulence case actually leads compared to the 10 km/s. The small region at the higher $\mathrm{R_{Fe II}}$ in the last panel ($\mathrm{Z = 10Z_{\odot}}$; $\mathrm{v_{turb}}$=10 km/s) is probably a result of some thermal instabilities in the models and the corresponding discontinuous change in the cloud structure as the grids of the parameters in the models are homogeneous. This blob is no more seen when the observational cut (based on Figure \ref{fig:param_space}) to the parameter space is added.

To cover well the whole observed optical plane, we need a whole range of black hole masses, Eddington ratios, cloud densities, metallicities and turbulent velocity. As in P18, the average sources in \citealt{shen11} catalog,  are well modeled with just solar abundance and a range of densities. But now,
if some turbulence, high cloud density and high metallicity is allowed, our model covers also the region of the very high values of  $\mathrm{R_{Fe II}}$ parameter, up to $\sim 5 - 6$, so now even extreme Fe II emitters can be reproduced. This was not achieved in P18, so the presence of the warm corona brought our model closer to the observed properties of the quasar sample.

\subsection{Comparison of the model with the data coverage of the optical plane}

\begin{figure*}
    \centering
    \includegraphics[scale=0.475]{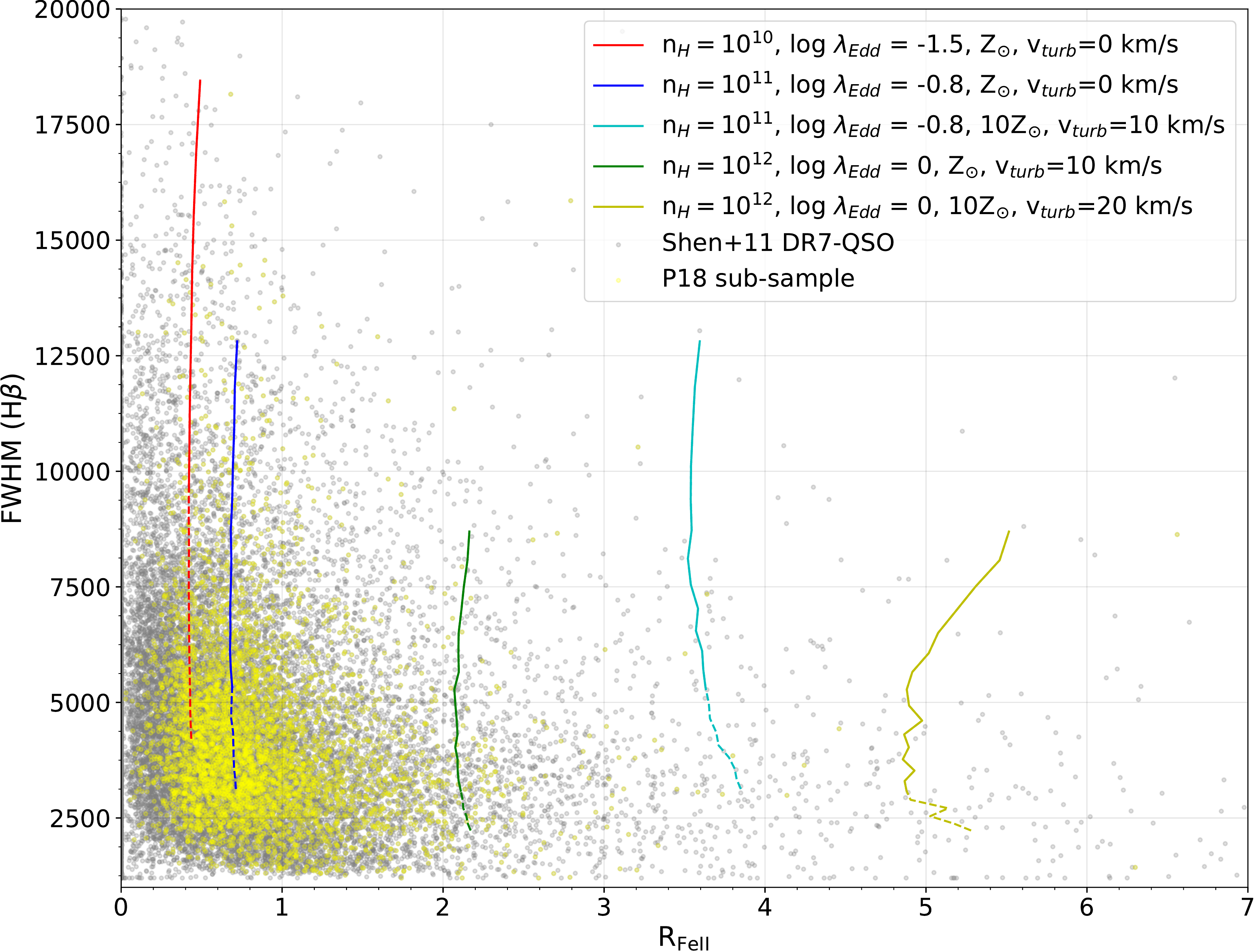}
    \caption{The coverage of the optical plane is shown with a broad range of parameters: mean cloud density (log n$\mathrm{_{H}}$ (in cm$^{-3}$) = 10 - 12), solar to 10 times solar abundances, 0 - 20 km s$^{-1}$ microturbulence, for a single density cloud at a high column density (log N$\mathrm{_H}$ (in cm$^{-2}$) = 24). The observational data points from the \citet{shen11} DR7-QSO catalogue, and a cleaner subset from \citet{panda18a}, are plotted to show the coverage of the modelled sequence.}
    \label{fig:opt_plane_data}
\end{figure*}

Previous sections show that with the present model we can represent even extreme Fe II emitters, but proper coverage of the optical plane requires also not to populate the part of the plane when real objects are 
not seen. With this aim, we selected the optimum but representative parameters examples and plotted them against the data points from the \citet{sh14} catalog, as we did in P18. The results are shown in Figure~\ref{fig:opt_plane_data}. Lines show only representative cases, and the spaces between the lines can be easily filled with models at intermediate parameter values than those presented in the plot. 

This optimum coverage requires the use of only low density, low metallicity, and low turbulence velocity clouds for low Eddington ratios, and a subsequent increase in the metallicity, density and turbulence with the Eddington ratio. Lowest density, metallicity and turbulence allows to recover objects located at the extreme left of the diagram. With higher density low Eddington ratio clouds, we would overpopulate the part of the diagram with high values of the FWHM at high values of $\mathrm{R_{Fe II}}$, where the real objects are rare. This is related to the fact that most effects (like changing metallicity or turbulent velocity; see Figure~\ref{fig:opt_plane_turbulence}) only weakly affect the line width. 

This selection is not entirely unique but it allows to better reproduce the observed Quasar Main Sequence. It also takes us back to some correlation between the metallicity and the Eddington ratio. While in Figure~\ref{fig:R_Fe} we showed that $\mathrm{R_{Fe II}}$ does not depend on the Eddington ratio $\lambda_{\mathrm{Edd}}$, now the dependence reappears as caused by the coupling between the $\lambda_{\mathrm{Edd}}$ and the cloud density and metallicity. Our model does not predict such coupling since for us all three quantities are free parameters. We cannot study quantitatively the coupling, as at that stage perhaps other couplings can be also suggested, and they would shrink the range of FWHM and  $\lambda_{\mathrm{Edd}}$ covered now by the model.

\subsection{Quasar Main Sequence in the UV }

Quasar Main Sequence is customarily studied in the optical plane, but similar study can be done in the UV plane. In this case the Mg II line at 2800 \AA~ has to be used instead of H$\beta$, and the Fe II optical emission has to be replaced with equally intense UV emission. Mg II and H$\beta$ both belong to Low Ionization Lines, as classified by \citet{collin88}, and thus should behave similarly. In \cite{Sniegowska18b} we  showed that the UV plane of the quasar main sequence based on Mg II line and Fe II (in the UV) emission indeed looks similar to the optical plane based on H$\beta$ line and Fe II (in the optical). 

\begin{figure*}
    \centering
    \includegraphics[scale=0.95]{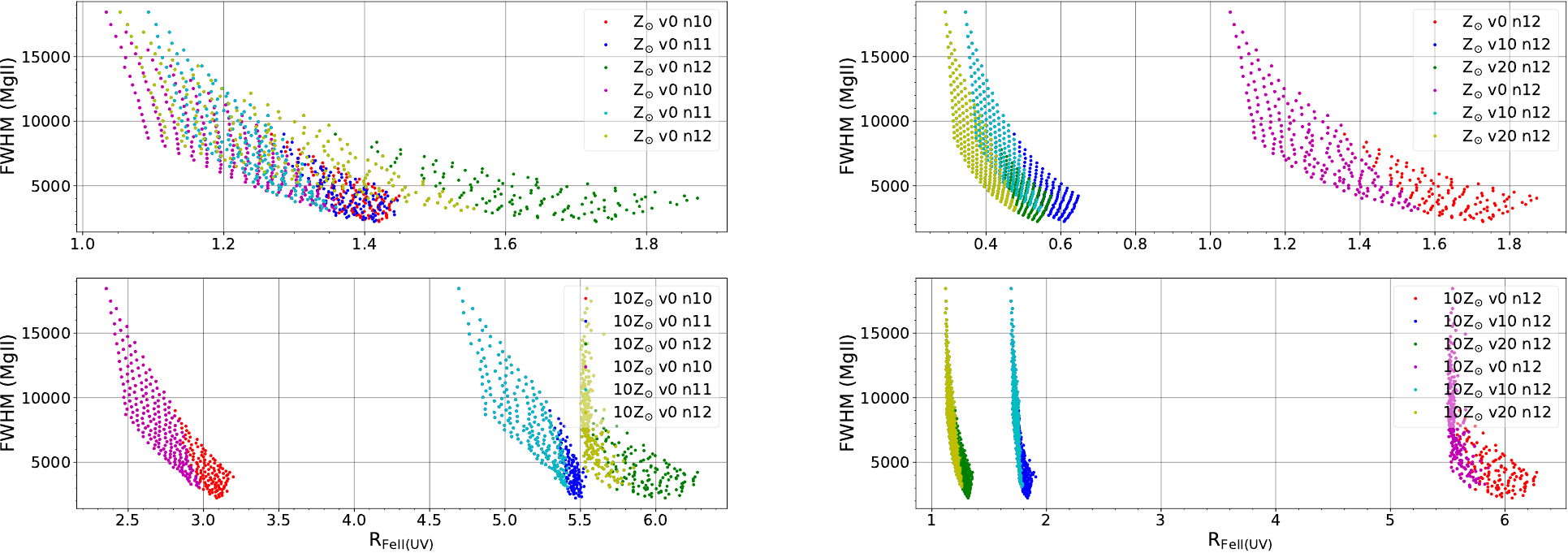}
    \caption{The coverage of the UV plane for two extreme cases of abundances (\textbf{Top panels}: at solar; and \textbf{bottom panels}: 10 times solar). \textbf{Left panels}: changing cloud density ($\mathrm{n_H = 10^{10}, 10^{11}\;and\;10^{12}}$ cm$^{-3}$) at zero turbulent velocity; \textbf{right panels}: changing turbulent velocities (0, 10 and 20 km/s) at constant cloud density ($\mathrm{n_H = 10^{12}\;cm^{-3}}$). The two cases are color-coded as in Fig.~\ref{fig:opt_plane_density}.}
    \label{fig:UV_plane}
\end{figure*}

We thus use our model to test whether it can also reproduce the Quasar Main Sequence in the UV plane. To do so, we used the warm corona models, used the same assumptions about the location of the BLR and cloud parameters, and with the use of the code CLOUDY we perform computations of the line intensities, constructing a similar main sequence diagram using the integrated Fe II emission strength in the UV. The range of the Fe II emission in the UV is considered to be within 2900 - 3050 \AA$\;$ (the redder side of the Fe II contamination in the Figure 5 from \citealt{kovacevic15}). This Fe II strength is derived by normalizing the integrated Fe II emission with the Mg II emission. This selection of the wavelength range is considered since at shorter wavelengths it is difficult to disentangle the blue wing of Mg II from Fe II contribution. 
\par 

Similar to the optical plane relations, we now show the predicted dependence of the FWHM of Mg II line on the parameter $\mathrm{R_{Fe} (UV)}$ measuring the relative strength of Mg II and UV Fe II (see Figure \ref{fig:UV_plane}). The results again depend on the adopted density, turbulence and metallicity, but the trends are not the same as in the case of the optical plane. There is for example large overlap between the plots for different densities or turbulent velocities, if solar metallicity is assumed.  The non-monotonic behaviour in the top panels of Figure \ref{fig:UV_plane} nearly goes away as the abundances are increased (from $\mathrm{Z = Z_{\odot}}$ to $\mathrm{Z = 10Z_{\odot}}$). But, similar to the case of the optical plane, Fe II emissivity is strongly enhanced when metallicity higher than solar is introduced, and when BLR clouds have higher densities. 

 The most significant difference between the optical and the UV plane is in fact in the dependence on the turbulence velocity. Now, an increase of the turbulence from 0 to 20 km s$^{-1}$ results in a significant reduction of the Fe II emissivity. 

The range of the FWHM is same as before since we did not introduce any correction for the location of the peak of the Mg II emission. Observationally, there are some indications that Mg II is located somewhat further than H$\beta$ since the Mg II lines are narrower by a factor 0.81, as shown in \citep{wang09}, but the difference is not large and it was not included in the present study. 

\begin{figure*}
    \centering
    \includegraphics[scale=0.475]{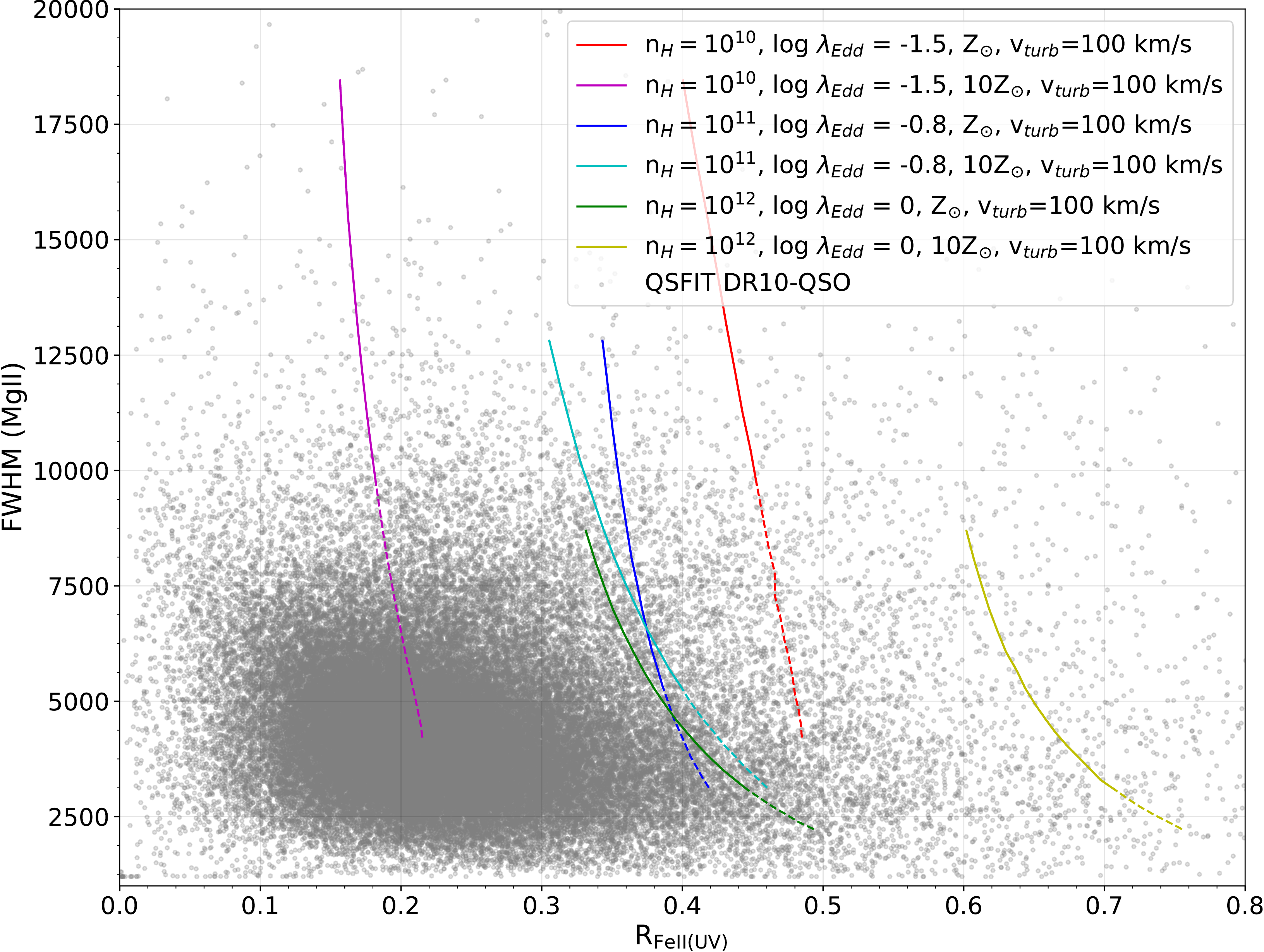}
    \caption{The coverage of the UV plane is shown with a broad range of parameters: mean cloud density (log n$\mathrm{_{H}}$ (in cm$^{-3}$) = 10 - 12), solar to 10 times solar abundances, at 100 km s$^{-1}$ microturbulence, for a single density cloud at a high column density (log N$\mathrm{_H}$ (in cm$^{-2}$) = 24). The observational data points from the QSFIT catalogue (\citealt{calderone17}) are plotted to show the coverage of the modelled sequence. 
    }
    \label{fig:UV_plane_data}
\end{figure*}

We now overplot a few sequences of selected models on top of the observational data. Data points and the models are shown in Figure~\ref{fig:UV_plane_data}. The data points, as in \citet{Sniegowska18b}, come from the QSFit catalog of \citet{calderone2017}.

Observationally, the UV quasar plane suggests somewhat narrower range of the FWHM than the optical plane, up to 9000 km s$^{-1}$ instead of 12000 km s$^{-1}$ in the optical plane for the same sample \citep{Sniegowska18b}. The observed values of $\mathrm{R_{Fe II}}$ (UV) are centered around $\sim 5$, and in the extreme cases extend up to 12 but in \citet{Sniegowska18b} we used a different wavelength range to measure the Fe II contribution (from 1250 \AA~ to 3090 \AA), much broader than in the present paper (2900 - 3050 \AA), so the direct comparison is not possible. We thus re-plot the data points from \citep{Sniegowska18b} re-scaling the values of $\mathrm{R_{Fe II}}$ (UV) by a factor of 0.085 obtained as a ratio between EW(Fe II) in the wavelength range used by QSFit \citep{calderone17} to the wavelength range used in the current study. This was done for the Fe II template of \citealt{vestegaard01} adopted by QSFit. 

We see from Figure \ref{fig:UV_plane_data} that the parameter range adjusted to fit the optical plane does not reproduce well the UV plane if we limit ourselves to the turbulent velocities between 0 and 20 km s$^{-1}$, as we did when modelling the optical plane. The predicted Fe II emissivities are much higher that seen in the data, with $\mathrm{R_{Fe II}}$ (UV) mostly above 0.4 while the data point concentrates at $\sim 0.2$. We may argue that the emission comes from a different region, but we prefer to check first if indeed it is necessary, and we reconsider the adopted parameter range for the turbulence.

In P18 we studied a much broader range of the turbulent velocities and we noticed that the dependence is not monotonic (see Fig.~5 in P18). The trend depends on the density but basically, for the optical Fe II emission the emissivity first rises with the turbulent velocity, and for values above 20 km s$^{-1}$ decreases again, so the case of high turbulence velocity is actually similar to the very low turbulence velocity. Thus, if we dramatically broaden the range of turbulent velocities in Fig. \ref{fig:opt_plane_data} (optical plane with data points) this coverage will not change. However, in the case of the UV plane, further increase of the turbulent velocity leads to further reduction of the Fe II emissivity, and with values of order of 100 km s$^{-1}$ we can now reach the center of the object population in the UV plane, and thus cover approximately the UV plane as well (see the line most to the left in Figure~\ref{fig:UV_plane_data}, still within the scheme of a single density, single distance model. 

\subsection{comparing Fe II strength in the optical and UV}

\begin{figure}
    \centering
    \includegraphics[scale=0.22]{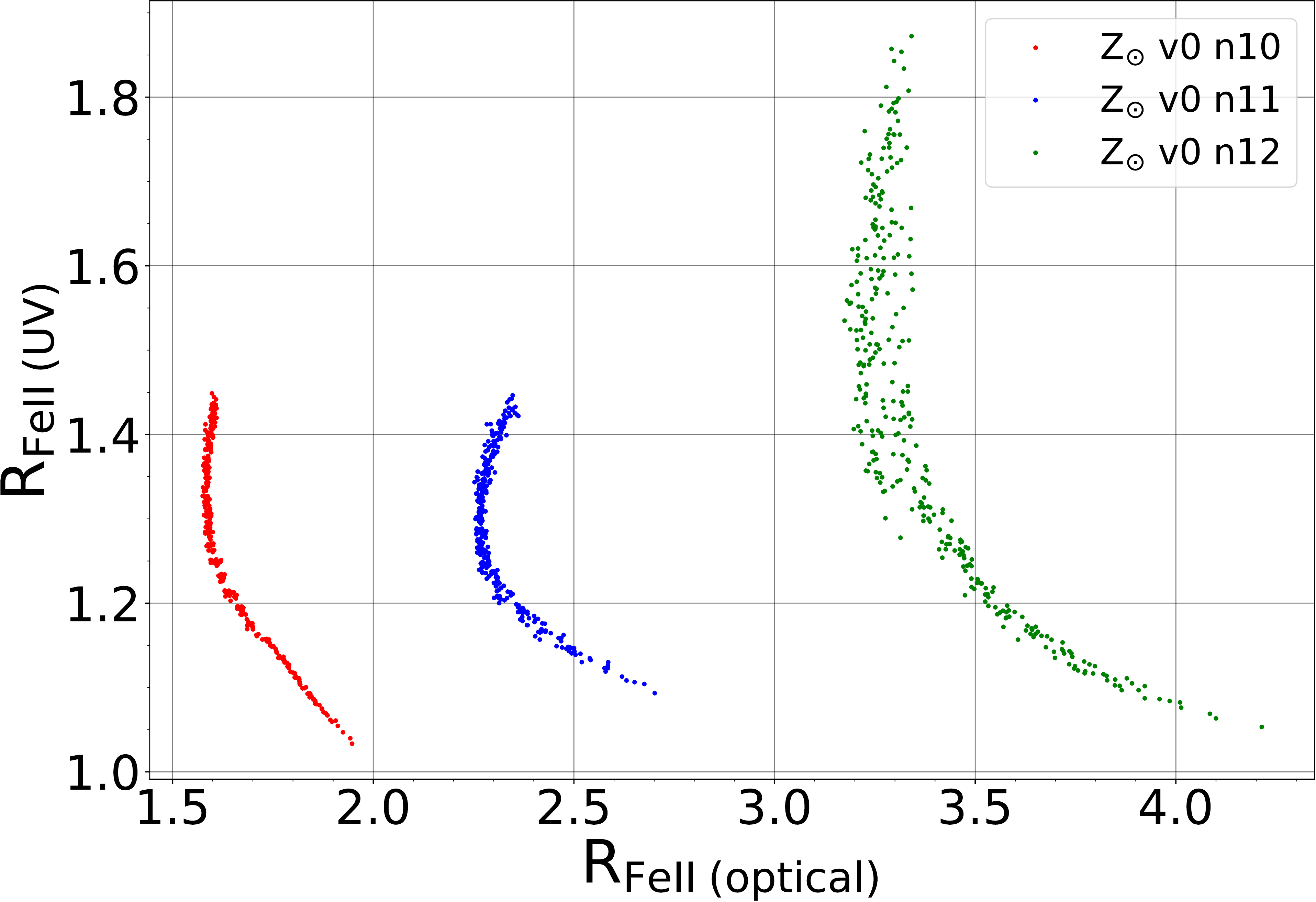}
    \caption{Fe II strength in optical vs UV. The optical Fe II range considered: 4434-4684 \AA, while in the UV it is: 2900-3050 \AA. The integrated Fe II emission are then normalised by broad H$\beta$ and Mg II, respectively, to get the corresponding Fe II strengths. The plots shown are for three cases of mean cloud density ($10^{10}, 10^{11}$ and $10^{12}$ cm$^{-3}$), at zero turbulence and solar metallicity.}
    \label{fig:R_Fe_opt_UV}
\end{figure}

We compare the Fe II strength obtained in the optical (integrated Fe II EW within 4434-4684 \AA$\;$ normalised by broad H$\beta$ EW) and UV (integrated Fe II EW within 2900-3050 \AA$\;$ normalised by broad Mg II EW). Figure \ref{fig:R_Fe_opt_UV} shows the dependence between the $\mathrm{R_{Fe II}}$ (optical) - $\mathrm{R_{Fe II}}$ (UV) for the WC model at solar abundances for three case of cloud densities (log $\mathrm{n_H}$ (in $\mathrm{cm^{-3}}$) = 10, 11, and 12) without turbulence. The exact values of the parameters depend predominantly on the set range of wavelength but the plot is an interesting illustration of the trends with the changing density. The two values are not proportional, as we might expect.  In some parameter range the relation predicts two distinct values of $\mathrm{R_{Fe II}}$ (UV) strengths for a given $\mathrm{R_{Fe II}}$ (optical). Therefore, on the basis of the model we would not expect a strong correlation between Fe II optical and UV emission, and indeed such correlation is not seen in the observational data \citep{kovacevic15}.  

\subsection{$R_{BLR}$ scaling}
\label{subsect:rblr}
\begin{figure*}
    \centering
    \includegraphics[scale=0.3]{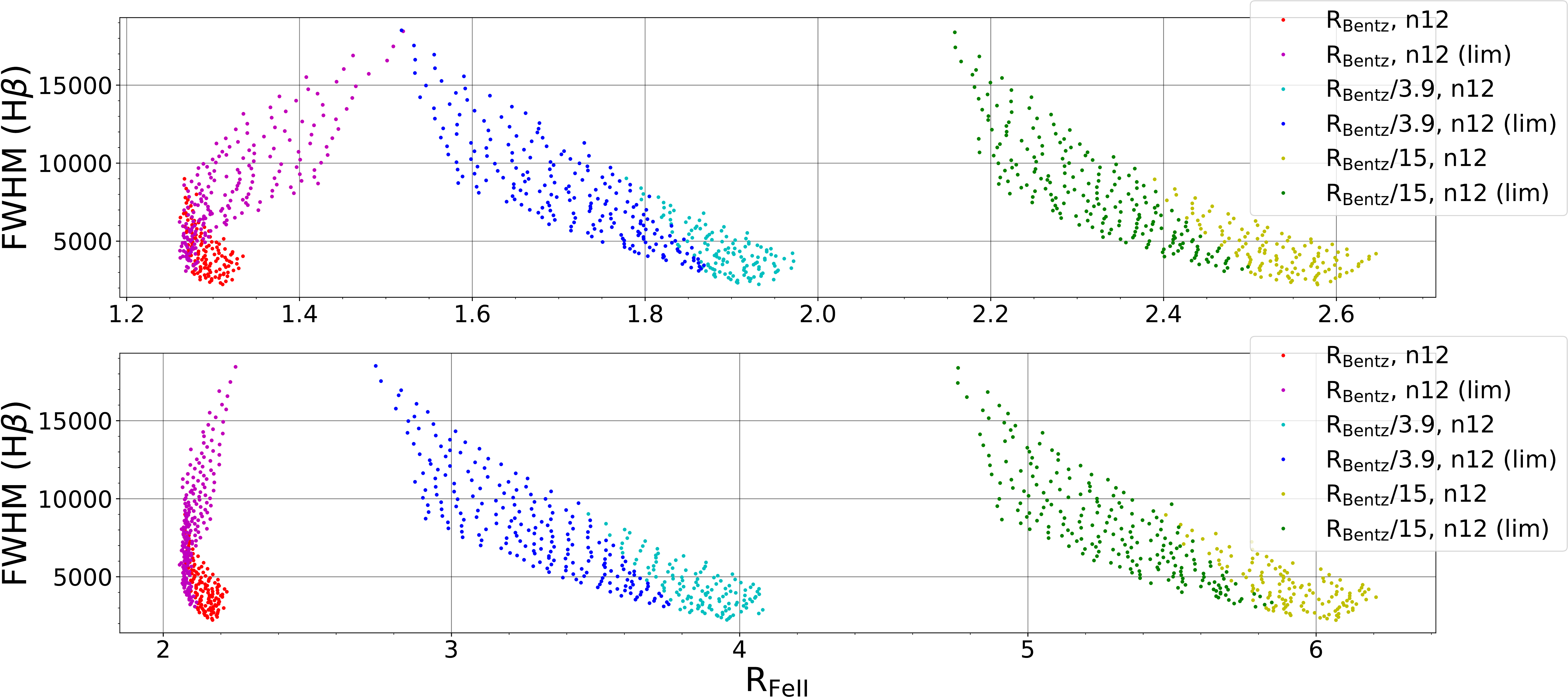}
    \caption{Fe II strength as a function of the size of the BLR. We show three cases of changing BLR size: at R$_{\mathrm{Bentz}}$ (\citealt{bentz13}), and at two other sizes scaled with respect to the R$_{\mathrm{Bentz}}$ (derived from the largest deviation, i.e. 15 days and its standard deviation ($\approx 3.9$ days) in the $\tau -\lambda L_{\lambda}$ relation (\citealt{grier17}). The panels are shown for the cloud's density $10^{12}$ cm$^{-3}$ at solar metallicity. The two cases are color-coded as Fig \ref{fig:opt_plane_density}. \textbf{Top panel}: at zero turbulence; \textbf{bottom panel}: at turbulence 10 km/s. The FWHMs for the R$_{\mathrm{Bentz}}$/3.9 and the R$_{\mathrm{Bentz}}$/15 cases have been scaled with a \textit{f} factor $\sqrt{3.9}$ and $\sqrt{15}$ respectively to recover the identical line widths as in the original R$_{\mathrm{Bentz}}$ case.}
    \label{fig:modified_radius}
\end{figure*}

Recent time-lag studies (\citealt{grier17}) show that the BLR size could be lower by a factor of 15 than the predicted sizes from the radius-luminosity relation (\citealt{bentz13}). In P18, we tested this effect of BLR clouds being closer than predicted by \cite{bentz13} which gave higher Fe II strengths. This can indeed be true for many of the quasars that have high Fe II contamination in their spectra. We have tried to incorporate this into the WC models. We tried to scale the BLR sizes to a lower limit of 15 times smaller going upto $\sqrt{15}$ smaller than the original BLR sizes obtained from the \cite{bentz13} relation. In Figure \ref{fig:modified_radius} we show the scaling of the $\mathrm{v_{FWHM}}-\mathrm{R_{Fe II}}$ relation when the radius of the BLR clouds is reduced by these factors. Like in the previous plots, we show both the full sequence and the limited range from observations. We use a cloud density of $\mathrm{n_H\;(in\;cm^{-3})} = 10^{12}$ at solar abundances for two cases of turbulence ($\mathrm{v_{turb}}$= 0 and 10 $\mathrm{km\;s^{-1}}$). As expected, the spread in the optical plane monotonically increases as the size gets smaller. Here we did not change the corresponding FWHM, since in this case the \cite{bentz13} relation does not apply. It may seem that lines should become broader as we move the BLR closer in but actually the change can be equally well absorbed by the virial factor, in general present in the mass-radius-velocity relation. We would need another method to locate the BLR, and to determine the line width, for a given black hole mass. For example, \citet{czerny2018} suggests returning to the size-luminosity relation based on bolometric flux while \cite{dupu2016a} argues that with the inclusion of the high-Eddington sources the scatter in the R-L relation shows a clear departure from the one-to-one relation in \citet{bentz13}. Discrimination between the two options is beyond the scope of the present paper since these short lags come from a relatively short campaign and need confirmation. However, a strong trend of increase in the Fe II is interesting.  

\section{Discussions}

The aim of the project was to test our understanding of the Quasar Main Sequence by an attempt to reproduce the distribution of the observational points in the optical plane with the theoretical model. The model assumed a grid in the black home masses, and Eddington ratios. The distance to the BLR was assumed using the radius-luminosity relation of \citet{bentz13}, which gave us the line width, and the emission line fluxes were calculated using the code CLOUDY v17.01 \citep{f17}, for a range of densities, metallicities and turbulent velocities. In P18 we were able to cover most of the region apart from the strongest Fe II emitters. In the present work we cover this region as well (see Fig.~\ref{fig:opt_plane_data}) since we now include the warm corona in modelling the object SED.

The presence of the warm corona affects significantly the Fe II and H$\beta$ emissivity. Without warm corona, the line ratio $\mathrm{R_{Fe II}}$ was sensitive to the Eddington ratio, while with warm corona this dependence disappeared. This means that high Eddington ratio sources can be found between strong Fe II emitters as well as between weak Fe II emitters, consistent with classical concept of NLS1 as high Eddington ratio sources, independently from the Fe II strengths. This would need further support from individual modeling of sources located in the left lower corner of the optical plane.

Good coverage of the whole plane requires a range of cloud densities, turbulence and metallicities since high density high metallicity clouds are more efficient Fe II emitters. However, simple increase of the metallicity led only to displacement of the modelled sequence rightwards, which was enough to cover the region of high Fe II emitters at low values of FWHM but at the same time over-predict the number of high FWHM emitters at that $\mathrm{R_{Fe II}}$ location. In order to cover the optical plane more precisely instead of too broadly, we need a coupling between these quantities. We see from the trends presented in Figure~\ref{fig:opt_plane_density} that, for a fixed density, the highest values of FWHM correspond to the highest black hole masses and the lowest Eddington ratios. If we postulate that density and/or metallicity rises with the Eddington ratio or decrease with the black hole mass then we could reproduce the coverage of the optical plane more precisely, populating the right part of the diagram mostly with high Eddington small mass sources. Such a trend has been noticed already (for both Eddington ratio and black hole mass) by \citet{sh14}. However, the procedure is not unique, for example rise in the density gives qualitatively similar effect to the rise of metallicity so we do not attempt to perform this exercise quantitatively.  

The current model still has problems with reproducing the lowest values of the line widths. This may be due to the use of a fixed virial vactor 1 connecting the black hole mass, BLR distance, and FWHM of the lines. If we adopt for example the virial factor of 1.3, as recently derived by the \citet{sturm2018} for spatially resolved BLR in 3C 273, the line width values become smaller by a factor of 14 \%. However, the virial factor likely depends on the FWHM itself, as argued by \citet{mejia18}, and the suggested virial factor range implies a possibility of line widths smaller even by a factor 2. Additionally, the problem of the line width range may be partially related to the viewing angle dependence of the line width, which is not yet included in our model.  

In the same way as we modeled the optical plane, we also modeled the UV plane observationally discussed by \citet{Sniegowska18b}. In this case H$\beta$ is replaced with Mg II line, and the optical Fe II emission with the UV Fe II emission. We performed the modelling using CLOUDY v17.01 and the whole methodology as before. The modeling was overall successful, we can recover the main trends for the following range of parameters: log n$\mathrm{_{H}}$ (in cm$^{-3}$) = 10 - 12, solar to 10 times solar abundances, 0 - 20 km s$^{-1}$ microturbulence, for a single density cloud at a high column density (log N$\mathrm{_H}$ (in cm$^{-2}$) = 24). The predicted line width were as before since in our method we assumed the same location for H$\beta$ and Mg II which is a good approximation.

In this model of the optical plane we use only a single cloud density and position as a representation of the whole BLR which we know is extended. We tested the dependence of the line luminosity on that mean radius, and we noticed a considerable rise in the relative Fe II optical luminosity. However, we cannot predict the net result of the broadening of the region since this would require arbitrary assumption about the cloud distribution as a function of the radius, which would provide us with the relative importance of the different BLR radii. 

\section{Conclusions}
We show that a simple model of H$\beta$ and Fe II production in the optical band, with minimum number of free parameters, is able to reproduce the observed coverage of the optical plane by quasars from \citet{shen11} catalog. The presence of the warm corona in the quasar SED is an important element, decreasing the dependence of the parameter $\mathrm{R_{Fe II}}$ on the Eddington ratio. The full coverage of the plane requires the presence of the sources with high metallicity although the central part of the quasar distribution is well recovered with solar metallicity. UV plane is not so well reconstructed, and the current model requires very high turbulence velocity. Further research of the UV plane coverage is clearly needed.
The key parameters behind the Quasar Main Sequence are black hole mass, Eddington ratio, cloud density and metallicity, and the two last quantities are are likely correlated to the first two. 
\section*{Acknowledgements}
The project was partially supported by the Polish Funding Agency National Science Centre, project 2015/17/B/ST9/03436/ (OPUS 9).
\software {CLOUDY v17.01 (\citealt{f17}); MATPLOTLIB (\citealt{hunter07})}
\bibliographystyle{aasjournal}
\bibliography{warm_corona}

\end{document}